\documentclass[aps,floatfix,prb,twocolumn]{revtex4}%
\usepackage{amsfonts}
\usepackage{amsmath}
\usepackage{amssymb}
\usepackage[final,dvips]{graphicx}%
\begin{document}
\title{Ground states of the Josephson vortex lattice in layered superconductors}
\author{A. E. Koshelev}
\affiliation{Materials Science Division, Argonne National Laboratory,Argonne, Illinois 60439}

\begin{abstract}
We consider the ground state configurations of the Josephson vortex
lattice in layered superconductors. Due to commensurability effects with
the layered structure, the lattice has multiple configurations, both
aligned with layers and rotated at finite angle. At low fields the lattice
switches between these configurations via first-order phase transitions.
These transitions become more frequent at smaller fields. With increasing
magnetic field a dilute lattice transforms first into a sheared dense
lattice. With further increase of field, the shear deformation smoothly
vanishes at a second-order phase transition.
\end{abstract}
\maketitle

A magnetic field applied parallel to the layers generates a lattice of
Josephson vortices with many unusual static and dynamic properties. An
important field scale is set by the anisotropy factor $\gamma$ and
interlayer periodicity $s$, $B_{cr}=\Phi_{0}/(2\pi\gamma s^{2})$ (in
Bi$_{2}$Sr$_{2}$CaCu$_{2}$O$_{8-\delta}$, $B_{cr}\sim 0.3-0.5$T). There
are two very different regimes depending on the strength of the magnetic
field. In the dilute lattice regime, $B_x\ll B_{cr}$, the nonlinear cores
of Josephson vortices are well separated and the distribution of currents
and fields is very similar to that in continuous anisotropic
superconductors. At high fields, $B_x>B_{cr}$, the dense lattice regime is
realized where the cores of the Josephson vortices overlap. In this regime
the Josephson vortices fill all layers homogeneously
\cite{BulaevskiiC1991}. This state is characterized by very weak
modulations of the in-plane and Josephson currents. In this proceedings we
consider the evolution of the ground state configuration with increasing
magnetic field.
%

\begin{figure}[ptb]
\centerline{\includegraphics[width=3in]{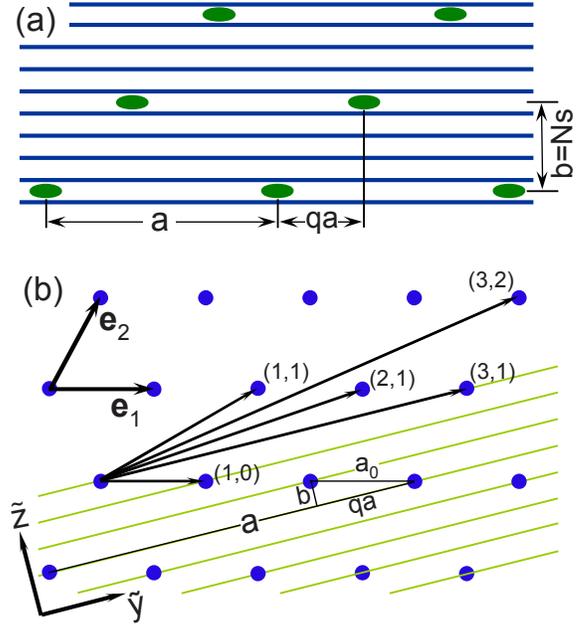}}\caption{ (a)
General Josephson vortex lattice and its parameters. (b) Orientation of
layered structure with respect to ideal lattice (in scaled coordinates).
Layered structure fits the ideal lattice only if it is oriented along one
of the crystallographic directions, which is characterized by two numbers
$(n,m)$, corresponding to expansion of the direction vector over the two
basic lattice vectors, $\mathbf{e}_{1}$ and $\mathbf{e}_{2}$. Several
possible directions are shown with the corresponding indices $(n,m)$. The
lower part of the figure illustrates the lattice parameters, $a$, $b$, and
$q$ for one possible orientation.}%
\label{fig:JVLcommens}%
\end{figure}

As the centers of the Josephson vortices must be located between the
layers, the layered structure plays crucial role in selection of the
ground-state lattice configurations. The Josephson vortex lattice is
commensurate with the layered structure only at the discrete set of
magnetic fields. At small magnetic fields, $B_x<\Phi_0/(2\pi\gamma s^2)$,
the Josephson vortex lattice can be described by anisotropic London model
(see, e.g., Refs.\ \onlinecite{Levitov1991,IvlevKP1990}). This model has a
simple scaling property: in scaled coordinates, $\tilde{z}=z/s$,
$\tilde{y}=y/\gamma s$, the energy becomes isotropic in
$\tilde{z}$-$\tilde{y}$ plane, which means that the ground state in these
coordinates corresponds to an ideal triangular lattice which is degenerate
with respect to rotations in this plane. In real coordinates these
rotations correspond to \textquotedblleft elliptic
rotations\textquotedblright. As a consequence, the family of commensurate
lattices includes lattices aligned with the layers, as well as misaligned
ones. To make a full classification of commensurate lattices we consider a
general lattice shown in Fig.\ \ref{fig:JVLcommens}a
\cite{Levitov1991,IvlevKP1990}. The lattice is characterized by three
parameters: in-plane period $a$, distance between vortex rows in $c$
direction $b=Ns$, and relative shift between the neighboring vortex rows
in $c$ direction $qa$. The lattice shape is characterized by the two
dimensionless parameters, $q$ and ratio $r=b/a$. The lattice parameters
are related to the in-plane magnetic field, $B_{x}$, as
$B_{x}=\Phi_{0}/(ab)$. Structures aligned with the layers correspond to
$q=1/2$. As the replacement $q\rightarrow1-q$ corresponds to the mirror
reflection with respect to x-z plane, every
structure with $q\neq1/2$ is double-degenerate.%

We start with classification of the lattices exactly commensurate with the
layered structure giving the set of commensurate fields (see also  Ref.\
\onlinecite{LagunaDB2000}). The analysis of commensurability conditions
can be done most conveniently in scaled coordinates
$(\tilde{y},\tilde{z})$. In these coordinates the ground-state
configuration corresponds to regular triangular lattice with period
$a_{0}=\sqrt{2 \Phi_{0}/\sqrt{3}\gamma s^2 B_{x}}$. It is convenient to
consider orientation of the layered structure with respect to this lattice
rather than the other way round. The layered structure fits this lattice
only if it runs along one of the crystallographic directions, see Fig.\
\ref{fig:JVLcommens}b. The direction $(n,m)$ is defined by the lattice
vector, $\mathbf{e}_{n,m}$, which
can be expanded over the two basic lattice vectors, $\mathbf{e}_{(n,m)}%
=n\mathbf{e}_{1}+m\mathbf{e}_{2}$. For nonequivalent directions $n$ and
$m$ must be relatively prime numbers. In particular, two aligned
configurations correspond to $(n,m)=(1,0)$ and $(1,1)$. Any such direction
corresponds to set of matching fields, $B_{(n,m)}(N)$, which can be found
by direct geometrical calculation\cite{LagunaDB2000}
\begin{equation}
B_{(n,m)}(N)=\frac{\sqrt{3}}{2}\frac{\Phi_{0}}{N^{2}\gamma s^{2}%
(n^{2}+nm+m^{2})}. \label{eq:CommensFields}%
\end{equation}
This result essentially rely on the London approximation, which implies a
very strong inequality $N\sqrt{n^{2}+nm+m^{2}}\gg 1$. Number of
vortex-free layers per unit cell is given by $N-1$. The case $N=1$
represents a special situation when all the layers are filled with
vortices and equivalent. It is interesting to note that even for dilute
lattice one can have the Josephson vortices in every layer ($N=1$) in the
case of high-order commensurability ($n,m\gg1$). In ideal situation the
lattice switches between different commensurate configuration via series
of first-order phase transitions. Number of competing states rapidly
increases with field decrease. In addition to giving general ground state,
these lattices describe multiple metastable states with unique
hierarchical properties studied in Refs.\
\onlinecite{IvlevKP1990,Levitov1991}.

Full analysis of the structure evolution requires energy consideration.
The London model does not not describe layered superconductor at high
fields. To obtain lattice structures in this region one has to consider
more general Lawrence-Doniach model. The transition between the
\emph{aligned} lattices have been studied within this model by Ichioka
\cite{Ichioka95}. However, our analysis shows that at many fields the true
ground state is not given by aligned lattice. If we limit ourself only to
aligned configurations, we reproduce results of Ref.\
\onlinecite{Ichioka95} at high anisotropies.

Within the Lawrence-Doniach model, at fields
$B_{x}\gg\Phi_{0}/(4\pi\lambda \lambda_{c})$ the lattice energy can be
represented as
\begin{equation}
f_{\mathrm{Jl}}=\frac{B_{x}^{2}}{8\pi}+\frac{B_{x}\Phi_{0}}{(4\pi)^{2}%
\lambda\lambda_{c}}u(N,q,h)\label{eq:LDLatEnTotal}%
\end{equation}
where $\lambda\equiv \lambda_{ab}$ and $\lambda_{c}$ are the components of
the London penetration depth, $h\equiv2\pi\gamma s^{2}B_{x}/\Phi_{0}$ is
the reduced magnetic field, and the reduced energy $u(N,q,h)$ is given by
\begin{align}
u(N,q,h)  & =\frac{1}{\pi}\sum_{n=1}^{N}\int d\tilde{y}\left(
\frac{1}{2}\left(
\frac{d\phi_{n}}{d\tilde{y}}\right)  ^{2}+\right.  \label{eq:LDEner}\\
& \left.  +1-\cos\left(  \phi_{n+1}-\phi_{n}-h\tilde{y}\right)  \right)
\nonumber
\end{align}
To match the London limit\cite{IvlevKP1990}, we write $u(N,q,h)$ in the
form
\begin{equation}
u(N,q,h)=\frac{1}{2}\ln\frac{1}{h}+1.432+G(N,q,h)\label{eq:LDEner-pres}%
\end{equation}
where the function $G(N,q,h)$ defined by this equation approaches the
London limit\cite{IvlevKP1990}, $G_{L}(r=N^{2}h/(2\pi),q)$, for
$h\rightarrow0$ with
\[
G_{L}(r,q)=\frac{\pi r}{6}+\sum_{l=1}^{\infty}\frac{1}{l}\frac{\exp\left(
-2\pi rl\right)  +\cos\left(  2\pi ql\right)  }{\cosh\left(  2\pi rl\right)
-\cos\left(  2\pi ql\right)  }-\frac{1}{2}\ln(2\pi r).\label{eq:gfun}%
\]
This function depends only on lattice shape. Its absolute minimum
corresponding to the triangular lattice is given by
$G_{L}(\sqrt{3}/2,1/2)=-0.4022$.

\begin{figure}[ptb]
\begin{center}
\includegraphics[width=3.29in]{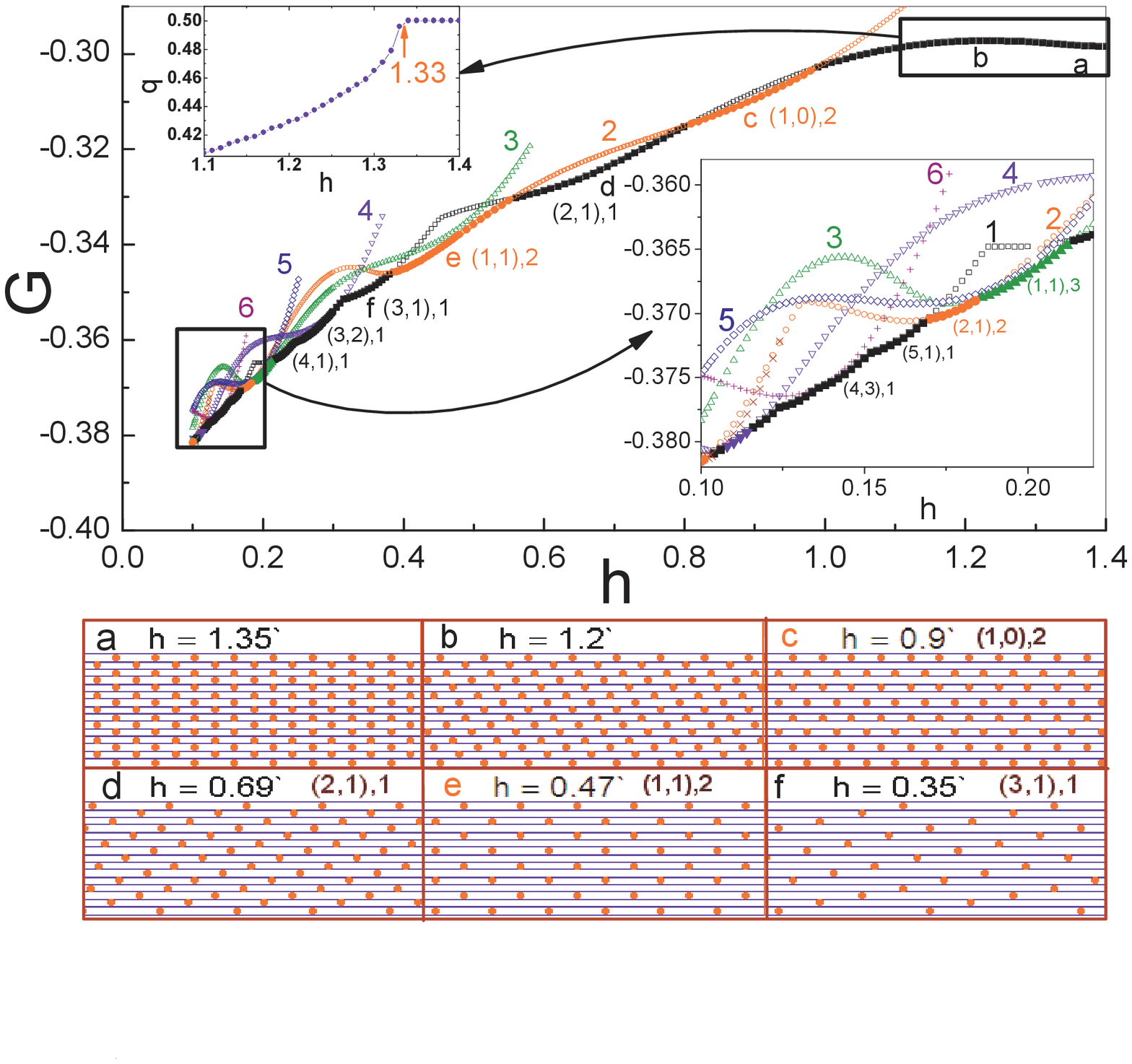}
\end{center}
\caption{The field dependence of the energy function $G(N,q,h)$ for the
ground state and competing states. Each branch corresponds to the minimum
of $G(N,q,h)$ with respect to $q$ at fixed $h$ and $N$. The curves are
marked by value of $N$. Lattice configurations in scaled coordinates are
shown at six marked fields. The left inset shows the field dependence of
the shift parameter $q$ near the structural transition of the dense
lattice. The right inset illustrates competition between different
configuration at smaller fields. The low-field local minima are marked by
the indices corresponding to commensurate configurations in the format
$(n,m),N$.}
\label{fig:G-h-Configs}%
\end{figure}

We explore the evolution of the ground-state configuration by direct
numerical minimization of the energy (\ref{eq:LDEner}) with respect to
lattice parameter $N$ and $q$ defined in Fig.\ \ref{fig:JVLcommens}, i.e.,
we computed reduced ground-state energy defined as
$G(h)\equiv\min_{N,q}[G(N,q,h)]$. The field dependence of the energy
function $G(N,q,h)$ is shown in Fig.\ \ref{fig:G-h-Configs} for the ground
state and competing states. Each curve corresponds to the minimum of
$G(N,q,h)$ with respect to $q$ at fixed $h$ and $N$ and it is marked by
the value of $N$. Every branch crossing corresponds to a first order phase
transition between different commensurate states. Below, we show the first
six lattice configurations which are realized with field decrease. At low
fields we specify for each of these ground-state configurations
corresponding indices $(n,m)$ and period $N$ in the format $(n,m),N$. At
small fields the local minima of branches occur at commensurate fields
corresponding to (\ref{eq:CommensFields}).

Let us review the lattice evolution with decreasing magnetic field. At
high fields an aligned dense Josephson vortex lattice is realized
(structure $a$). This lattice becomes unstable at $h\approx 1.33$. Below
this value, shear deformation develops corresponding to decrease of $q$
below $1/2$ (see structure $b$). The field dependence of $q$ near the
transition point is shown in the left inset of Fig.\
\ref{fig:G-h-Configs}. At $h\approx 0.99$ this sheared dense lattice is
replaced by the aligned lattice with the period $N=2$ (structure $c$). At
$h\approx 0.8$ this structure transforms back into the misaligned lattice
with period $N=1$ (structure $d$). At smaller fields, many lattice
configurations compete for the ground state, as one can see more clear in
the right inset. As a consequence transitions become more and more
frequent at smaller field. At several fields (e.g., at
$h\approx,0.19,0.137,0.105\ldots$) one or more lattice configurations have
energies very close to the ground-state energy. We also note that there
are several extended field ranges where in the ground state all layers
homogeneously filled with vortices ($N=1$) even in the region of dilute
vortex lattice, e.g., for $0.115<h<0.17$, $0.21<h<0.38$. The layered
structure favors such states. We also found that the layered structure
does not favor aligned structures with indices $(n,m)=(1,0)$. Such
structures do not realize in ground state for $2<N<7$.

In conclusions, we explored ground states of the Josephson vortex lattice
in layered superconductors. With decreasing field a dense lattice
transforms into a dilute lattice via an intermediate sheared dense lattice
state. After that the lattice goes through sequence of states with
different orientations and c-axis periods separated by first-order phase
transitions. At low fields many lattice configurations compete for ground
state and phase transitions become very frequent. As the energy
differences between different states induced by the layered structure
become tiny at small fields, external factors, such as interactions with
boundaries or with correlated disorder, may play role in selection of
ground states in real samples. Frequently such external interactions favor
configurations aligned with the layered structure.

I am grateful to X.\ Xu for finding error in one of the figures. This work
was supported by the U.\ S.\ DOE, Office of Science, under contract \#
W-31-109-ENG-38.


\end{document}